\documentclass[graybox,footinfo]{svmult}

\smartqed
\usepackage{mathptmx}       
\usepackage{helvet}         
\usepackage{courier}        
\usepackage{type1cm}        
\usepackage{graphicx}       

\usepackage{array,colortbl}
\usepackage{amsmath,amsfonts,amssymb,bm} 

\newcommand{\R}{\mathbb{R}} 
\newcommand{\bP}{\mathbb{P}} 
\newcommand{\N}{\mathbb{N}} 

\newcommand{\M}{{\mathcal M}}

\newcommand{\Exp}    {\operatorname{E}}
\newcommand{\V}    {\operatorname{Var}}
\newcommand{\err}  {\operatorname{error}}
\newcommand{\cost} {\operatorname{cost}}

\newtheorem{rem}{Remark}

\usepackage{microtype} 

\usepackage[colorlinks=true,linkcolor=black,citecolor=black,urlcolor=black]{hyperref}
\usepackage{bookmark}
\makeatletter 
  \providecommand*{\toclevel@author}{999}
  \providecommand*{\toclevel@title}{0}
\makeatother

\begin{document}

\title*{Variance Reduced Multilevel Monte Carlo Path Simulation: going beyond the complexity $\varepsilon^{-2}$ }
\titlerunning{Variance Reduced MLMC Path Simulation: going beyond the complexity $\varepsilon^{-2}$}
\author{Denis Belomestny \and Tigran Nagapetyan}
\institute{
Denis Belomestny
\at Duisburg-Essen University, Duisburg, Germany
\\ \email{belomestny@uni-due.de}
\and
Tigran Nagapetyan
\at Department of Statistics, University of Oxford
24--29 St Giles',
Oxford OX1 3LB,
England,\\ \email{nagapetyan@stats.ox.ac.uk}
}
\maketitle

\abstract{
In this paper a novel modification of the Multilevel Monte Carlo approach (MLMC), allowing for further significant complexity reduction, is proposed. The idea of the modification is to use a specifically designed control variate in the first level of  MLMC. We show that under a proper choice of the control variate, one can reduce the complexity order of the modified MLMC algorithm down to $\varepsilon^{-2+\delta}$ for any  $\delta\in [0,1)$ with $\varepsilon$ being the precision to be achieved. 
These theoretical results are illustrated by several numerical examples. }
\section{Introduction}
The multilevel path simulation method introduced in Giles \cite{giles2008multilevel} has recently gained  a lot of popularity as a complexity reduction tool. The main advantage of  the MLMC methodology is that it can be straightforwardly  applied to various situations and 
requires almost no prior knowledge on the path generating process. Any multilevel Monte Carlo (MLMC) algorithm uses a number of levels of resolution, $l=0,1,\ldots, L,$ with $l=0$ being the coarsest, and $l=L$ being the finest. In the context of SDE path simulation problem on the interval $[0,T]$, level $0$ corresponds to one time step $\Delta_0=T,$ whereas the level $L$  is related to $2^L$ uniform timesteps $\Delta_L=2^{-L}\cdot T.$ 
\par
The problem of interest in this paper is to estimate the quantity $a := \Exp(f(X_T))\in\R$ for a real-valued functional $f$ with a given accuracy $\varepsilon$, where $(X_t)_{t\in [0,T]}$ is a stochastic process, which can't sampled directly. We compare different Monte Carlo algorithms with their cost error relations, where by the cost and the error of the Monte Carlo algorithm $\M$ we understand $\cost (\M) = \Exp (\text{\# operations and random number calls})$ and  $\err(\M)=\left(\Exp(a-\M)^2\right)^{1/2},$ respectively.
We say that a sequence of the Monte Carlo algorithms $\M_n$
with $\lim\limits_{n \to \infty} \err (\M_n) = 0$
achieves the order of convergence $\gamma > 0$ if there exists  $c > 0$ and $\eta\in\R$, such that
$$\forall\, n \in \N:\ 
\cost (\M_n) \leq c \cdot \bigl(\err(\M_n)\bigr)^{-\gamma}\cdot\left(-\log\err(\M_n)\right)^{\eta}.
$$
Note, that  the smaller $\gamma$ is, the better is the performance of the Monte Carlo algorithm. In our setting we can't sample $X_T$ directly, but we can sample $X_{l,T},$ where we assume that
$$\lim\limits_{l \to \infty} \cost(f(X_{l,T})) = \infty,\  \Exp f(X_{l,T})\to \Exp f(X_T).$$
The multilevel path simulation approach consists in first writing the expectation of the finest approximation $\Exp[f(X_{L,T})]$ as a telescopic sum
\begin{eqnarray}
\label{telescop_o}
\Exp[f(X_{L,T})]=\Exp[f(X_{0,T})]+\sum_{l=1}^L \Exp[f(X_{l,T})-f(X_{{l-1},T})]
\end{eqnarray}
and then applying  Monte Carlo  to estimate each expectation in this sum.
One important prerequisite for MLMC to work is that $X_{l,T}$ and $X_{{l-1},T}$ are coupled in some way and this can be achieved by using the same discretized trajectories of the underlying diffusion process to construct  the consecutive approximations $X_{l,T}$ and $X_{{l-1},T}.$ The degree of coupling is usually measured in terms of the variance $\V[f(X_{l,T})-f(X_{l-1,T})]$. 
It is shown in Giles \cite{giles2008multilevel}, that under the conditions: 
\begin{eqnarray}
\label{ml_ass}
\bigl|\Exp[f(X_{L,T})]-\Exp[f(X_T)]\bigr|\leq c_1\Delta_L^{\alpha},\quad \V\left[f(X_{l,T})-f(X_{{l-1},T})\right]\leq c_2 \Delta_l^{\beta}, 
\end{eqnarray}
with some $\alpha\geq 1/2,$ $\beta>0,$ $c_1>0$, $c_2>0$ and with cost of sampling $f(X_{l,T})$ bounded by $c_3\cdot \Delta_l^{-1}$, the computational complexity of the resulting multilevel estimate needed to achieve the accuracy $\varepsilon$ (in terms of RMSE)  is proportional to 
\begin{eqnarray}
\mathcal{C}\asymp
\begin{cases}
\varepsilon^{-2}, & \beta>1, \\
\varepsilon^{-2}\log^2(\varepsilon), & \beta=1, \\
\varepsilon^{-2-(1-\beta)/\alpha}, & 0<\beta <1.
\end{cases}
\label{MLMC:compl:Giles}
\end{eqnarray}
This is a significant improvement over the classical standard Monte Carlo approach, which has  complexity $\varepsilon^{-2-1/\alpha}$.
The above asymptotic  estimates however show that reduction of complexity beyond the order $\varepsilon^{-2}$ is not possible, doesn't matter how large is $\beta>1$. Moreover, there is an issue about achieving $\beta>1$ in higher dimensions with an implementable algorithm, which has bees successfully resolved in \cite{gs14} under certain regularity assumptions. This fact motivates a question on  existence of algorithms  with complexity order of order $\varepsilon^{-2+\delta},$ where $\delta>0$.  This is our main concern, and here we propose a modification of the original MLMC algorithm which makes further complexity reduction possible. 
Let us note that existence of such modification does not contradict the general lower bound in  \cite{creutzig2009infinite}, as 
the authors in   \cite{creutzig2009infinite} consider the case of general path dependent functionals of $(X_t)_{t\in [0,T]}$ and we study here  functionals of the form $f(X_T)$ under some additional smoothness assumption on $f.$ In this context let us mention the work \cite{muller1324deterministic},  where a deterministic quadrature rule  based on the distribution of a simplified weak Ito-Taylor step is proposed. In fact, the algorithm presented in \cite{muller1324deterministic}  also provides  complexity rates better than ones of the MLMC  algorithm, but its application is limited to one-dimensional case.
\par
The plan of the paper is as follows. The main idea of the variance reduced  MLMC approach  is introduced in \eqref{main_idea}. Section~\ref{control} is devoted to the construction of control variate.
\section{Variance Reduced MLMC}
\label{main_idea}
Fix some $0<L_0<L$ and consider a random variable $M_{L_0}$ with $\Exp\bigl[M_{L_0}\bigr]=0,$ then
\[
\mathrm{E}\bigl[f(X_{L,T})\bigr]=\mathrm{E}\bigl[f(X_{L_0, T})-M_{L_0}\bigr]+\sum\limits _{l=L_0+1}^{L}\mathrm{E}\bigl[f(X_{l,T})-f(X_{l-1,T})\bigr].
\]
As opposite to the representation \eqref{telescop_o}, we start the telescopic sum not at the roughest approximation  $\Delta_0=T$, but at some intermediate one corresponding to $\Delta_{L_0}$. Moreover, at level zero  we subtract a zero mean random variable $M_{L_0},$ which can be viewed as a control variate.  By fixing a vector of natural numbers $\mathbf{n}=(n_{L_0},\ldots,n_L)\in \mathbb{N}_0^{L-L_0+1},$ we can construct a modified multilevel Monte Carlo estimate for $Y=\mathrm{E}\bigl[f(X_{L,T})\bigr]$ via
\begin{eqnarray*}
\widehat Y\doteq\frac{1}{n_{L_0}}\sum_{i=1}^{n_{L_0}}\bigl[f(X^{(i)}_{L_0,T})-M^{(i)}_{L_0}\bigr]+\sum\limits _{l=L_0+1}^{L}\frac{1}{n_{l}}\sum_{i=1}^{n_{l}}\bigl[f(X^{(i)}_{l,T})-f(X^{(i)}_{{l-1},T})\bigr],
\end{eqnarray*}
where all pairs $\Bigl(X^{(i)}_{{l-1}},X^{(i)}_{l,T}\Bigr) $ are independent.
Obviously $\Exp\bigl[\widehat Y\bigr]=\mathrm{E}\bigl[f(X_{L,T})\bigr]$ and
$$\V\bigl[\widehat Y\bigr]\lesssim \frac{1}{n_{L_0}}\V\bigl[f(X_{L_0,T})-M_{L_0}\bigr]+\sum\limits _{l=L_0+1}^{L}n_l^{-1}\Delta_l^\beta,$$
where \(\lesssim\) stands for inequality up to a some constant not depending on \(L\) and \(\mathbf{n}\), provided the assumption \eqref{ml_ass} is fulfilled and $f$ is Lipschitz continuous. So we have for the mean square error of $\widehat Y,$
\begin{eqnarray}
\label{eq:MSE}
\Exp\bigl[|\widehat Y-\Exp \left[f(X_{T})\right]|^2\bigr]\lesssim \Delta_L^{2\alpha}+\frac{1}{n_{L_0}}\V\bigl[f(X_{L_0,T})-M_{L_0}\bigr]+\sum\limits _{l=L_0+1}^{L}n_l^{-1}\Delta_l^\beta.
\end{eqnarray}
Note that the cost of calculating $\sum\limits _{l=L_0+1}^{L}\frac{1}{n_{l}}\sum_{i=1}^{n_{l}}\bigl[f(X^{(i)}_{l,T})-f(X^{(i)}_{{l-1},T})\bigr]$ needs to be proportional to $\Delta_{L_0}^{\beta-1}\cdot\varepsilon^{-2}$ under conditions \eqref{ml_ass} with $\beta>1$. So we need to construct the control variate $M_{L_0}$ and calculate $\frac{1}{n_{L_0}}\sum_{i=1}^{n_{L_0}}\bigl[f(X^{(i)}_{L_0,T})-M^{(i)}_{L_0}\bigr]$ with the cost lower than $\varepsilon^{-2}$. We will measure the complexity of building and using the control variate $M_{L_0}$ in terms of $\Delta_{L_0}$. Moreover, we will assume, that the control variate $M_{L_0}$ satisfies 
$$\V\bigl[f(X_{L_0,T})-M_{L_0}\bigr]\lesssim \Delta_{L_0}^\mu,$$
with some $\mu>0$.  Our assumptions can be formalised as follows for certain $M>1$ and $\Delta_{\ell}=M^{-\ell}$.
\begin{align}
&\text{Sampling cost:}          & \cost\bigl(f(X_{l,T}),f(X_{{l-1},T}) \bigr)\leq c \cdot M^l
\label{as:cost} \\
&\text{Weak convergence: }      & \left| \Exp(f(X_T)) - \Exp ( f(X_{l,T}) ) \right|\leq c \cdot M^{-l \cdot \alpha},\ \alpha\ge1
\label{as:weak}\\
&\text{Degree of coupling:} & \Exp\left[ ( f(X_{l,T}) - f(X_{l-1,T}) )^2\right] \leq c \cdot M^{- l \cdot \beta},\ \beta>1                                                       \label{as:beta}\\
&\text{CV construction cost:}   & \cost_c\bigl(M_{L_0}\bigr) \leq c\cdot\Delta_{L_0}^{-\mu_1},\ \mu_1\ge0                                                                   \label{as:CVcon}\\
&\text{CV usage cost:}          & \cost_u\bigl(M_{L_0}\bigr) \leq c\cdot\Delta_{L_0}^{-\mu_2},\ \mu_1\ge\mu_2\ge0                                                        \label{as:CVuse}\\
&\text{CV effectiveness:}       & \V\bigl[f(X_{L_0,T})-M_{L_0}\bigr]\le c\cdot\Delta_{L_0}^{\mu_3}, \mu_3>\max(1,\mu_2)
\label{as:eff}
\end{align}
The above assumptions lead  to the following complexity theorem.
\begin{theorem}
\label{mlmc:cv:cost:thm}
Under the assumptions in \eqref{as:cost}-\eqref{as:eff}, we have with $\eta=\min(\beta-1,\mu_3-\max(\mu_2,1)),$ the overall cost of the variance reduced MLMC algorithm is proportional to
$$\mathcal{C}^\varepsilon_{\mu_1,\mu_2,\mu_3,\beta}= \varepsilon^{-\max\left(2-\frac{2\eta}{\eta+\max(\mu_1,1)},\frac{1}{\alpha}\right)},\quad \varepsilon\to 0.$$
\end{theorem}
\begin{proof}
It's easy to see, that with $\beta>1$ we have the overall cost proportional to 
\begin{gather*}
\max\left(\Delta_{L_0}^{\beta-1}\cdot\varepsilon^{-2},\varepsilon^{-\frac{1}{\alpha}}\right)+\Delta_{L_0}^{-\mu_1} + \max\left(\varepsilon^{-2}\cdot \Delta_{L_0}^{\mu_3},1\right)\cdot\left(\Delta_{L_0}^{-1}+\Delta_{L_0}^{-\mu_2}\right) 
\end{gather*}
or equivalently
\begin{multline*}
\Delta_{L_0}^{\beta-1}\cdot\varepsilon^{-2} + \varepsilon^{-\frac{1}{\alpha}}+\Delta_{L_0}^{-\mu_1}+\Delta_{L_0}^{-1}
+\Delta_{L_0}^{-\mu_2}+\varepsilon^{-2}\cdot \Delta_{L_0}^{\mu_3-\max(\mu_2,1)}\asymp\\
\Delta_{L_0}^{\eta}\cdot\varepsilon^{-2} + \varepsilon^{-\frac{1}{\alpha}}+\Delta_{L_0}^{-\max(\mu_1,\mu_2,1)}\Rightarrow \Delta_{L_0}\asymp \varepsilon^{\frac{2}{\eta+\max(\mu_1,1)}}.
\end{multline*}
Hence the overall cost is bounded by
$$
\mathcal{C}^\varepsilon_{\mu_1,\mu_2,\mu_3,\beta}=\varepsilon^{-\max\left(2-\frac{2\eta}{\eta+\max(\mu_1,1)},\frac{1}{\alpha}\right)}.
$$
\end{proof}
\begin{rem}
We see that, provided $\alpha>\frac12$, it doesn't matter how difficult it is to construct the control variate \(M_{L_0}\) or to use it (i.e. how large is \(\max(\mu_1,\mu_2,1)\)), if the goal is to get the complexity rate less than $\varepsilon^{-2}$ (i.e. to get \(\eta>0\)); what matters only is that  the efficiency of the control variate \(M_{L_0}\) is sufficiently large (i.e. $\mu_3>\max\{1,\mu_2\}$).
\end{rem}
\section{Construction of control variates for SDE}
\label{control}
In this section we are going to present a method of constructing control variates satisfying the assumptions \eqref{as:cost}-\eqref{as:eff}.
Let \(T>0\) be a fixed time horizon.
Consider a $d$-dimensional diffusion process
$(X_t;t\in [0,T])$
defined by the It\^o stochastic differential equation
\begin{align}
\label{x_sde}
dX_t
=\mu(X_t)\,dt
+\sigma(X_t)\,dW_{t},
\quad X_{0}=x
\end{align}
for continuous functions
\(\mu\colon\mathbb{R}^d\to\mathbb{R}^d\)
and
\(\sigma\colon\mathbb{R}^d
\to\mathbb{R}^{d\times m}\),
where \((W_t\in\mathbb{R}^m;t\in [0,T])\)
is a standard \(m\)-dimensional Brownian motion.
The coefficients $\mu$ and $\sigma$
are assumed to be such that there exists unique strong solution for \eqref{x_sde}.

\subsection{Some observations in one-dimensional case}
Our construction of the control variate will be connected to the Wiener Chaos decomposition (see \cite{nualart2006malliavin}  for a detailed exposition). Let $(\phi_i)_{i\geq 1}$ be an orthonormal basis in $L^2(0,T).$ The Wiener chaos of order $p\in \mathbb{N}$ is the $L^2$-closure of the vector field spanned by
\begin{eqnarray*}
\left\{\prod_{i\geq 1}  H_{p_i}\left(\int_0^T \phi_i(s)\,dW_s\right):\,\sum_{i\geq 1} p_i=p\right\},
\end{eqnarray*}
where $H_p$ is the Hermite polynomial of order $p$ given by the formula
\begin{eqnarray*}
H_p(x)\doteq\frac{(-1)^p}{\sqrt{p!}}e^{x^2/2}\frac{d^p}{dx^p}e^{-x^2/2}, \quad p\in \mathbb{N}_0.
\end{eqnarray*}
It is well known that $(H_p)_{p\geq 0}$ is a sequence of orthogonal polynomials in $L^2(\mathbb{R},\mu),$ where $\mu$ stands for centered Gaussian measure. 
Every square integrable random variable $F,$ measurable with respect to $\mathcal{F}_T,$ admits the  decomposition 
\begin{eqnarray}
\label{wiener_chaos}
F=\Exp[F]+\sum_{k\geq 1} \sum_{|p|=k} c_{p} \prod_{i\geq 1} H_{p_i}\left(\int_0^T \phi_i(s)\,dW_s\right)
\end{eqnarray}
with $p=(p_1,\ldots,p_k,\ldots)\in \mathbb{N}^{\mathbb{N}}$ and $|p|=\sum_{i\geq 1} p_i.$ Taking into account the orthogonality of Hermite polynomials, we derive an expression for the coefficients $c_p:$
\begin{eqnarray*}
c_p=\Exp\left[F\times \prod_{i\geq 1} H_{p_i}\left(\int_0^T \phi_i(s)\,dW_s\right)\right].
\end{eqnarray*}
In the situation where $F=f(X_{\Delta,T})$ and $X_{\Delta,T}$  comes from a discretisation of \eqref{x_sde} with a time step $\Delta=T/J$ for some  $J\in \mathbb{N},$ it is natural to take $\phi_i(t)\doteq\mathbb{I}\bigl(t\in ](i-1)\Delta,i\Delta]\bigr)/\sqrt{\Delta},$ $i=1,\ldots, J.$  If $X_{\Delta,T}$ is measurable with respect to $\mathcal{G}_J\doteq \sigma(\Delta_{1} W,\ldots,\Delta_{J} W)$ with $\Delta_{i} W\doteq W_{i\Delta}-W_{(i-1)\Delta}$ and $f(X_{\Delta,T})\in L^2(\mathcal{G}_J,\bP),$  then we obtain the decomposition
\begin{eqnarray}
\label{wiener_chaos_l}
f(X_{\Delta,T})=\Exp[f(X_{\Delta,T})]+\sum_{k\geq 1} \sum_{|p|=k} c_{p} \prod_{i=1}^{J} H_{p_i}(\Delta_{i} W/\sqrt{\Delta})
\end{eqnarray}
with $p=(p_1,\ldots,p_{J})\in \mathbb{N}_0^{J}.$ The above measurability assumption means that the approximation $X_{\Delta,T}$ involves only  uniformly-spaced discrete Brownian increments. This is, for example, the case for the Euler scheme and the Milstein scheme under the commutativity condition. Furthermore, 
Giles and Szpruch \cite{giles2012antithetic} constructed a coupled Milstein scheme that fulfils both  the above measurability assumption  and the condition  \eqref{ml_ass} with $\beta>1.$ Let us further analyse the decomposition  \eqref{wiener_chaos_l}. First note that the coefficients in \eqref{wiener_chaos_l} can be computed via
\begin{eqnarray*}
c_p=\Exp\left[f(X_{\Delta,T})\times \prod_{i=1}^{J} H_{p_i}\bigl(\Delta_{i} W/\sqrt{\Delta}\bigr)\right].
\end{eqnarray*}
So now we can consider a control variate of the form
\begin{eqnarray*}
M_{K,\Delta}\doteq\sum_{k=1}^K \sum_{|p|=k} c_{p} \prod_{i=1}^{J} H_{p_i}\bigl(\Delta_{i} W/\sqrt{\Delta}\bigr).
\end{eqnarray*}
Note that in order to compute all coefficients appearing in $M_{K,\Delta}$ we need $O(J^K)$ operations, which is unfeasible. We overcome this issue in the next section, where we suggest another  representation.

\subsection{Control variate construction in multidimensional case}
Let $d,J\in \mathbb{N}$, 
let $(\Omega,\mathcal{F},\bP,(\mathcal{F}_t)_{t\in [0,T]})$ be a filtered probability
space and let $W\colon [0,T]\times \Omega \rightarrow \R^m$
be a standard $(\mathcal{F}_t)_{t\in [0,T]}$-Brownian motion. For $J\in \mathbb{N}$ and 
$ j\in \{1,\ldots,J\}$ we define 
$ \Delta_{j} W = W_{j\Delta} - W_{(j-1)\Delta}$, where $\Delta = T/J$, and by $W^i$ we denote the $i$-th component of the vector. 
Let 
$\Phi_{\delta} \colon \R^{d+m}\to \R^d$
be measurable and suppose that for all $s,t\in [0,T]$ 
satisfying $s<t$ and $\delta\in[0,\infty)$,
there exists a constant $C\in [0,\infty)$ not depending on \(\delta\) and \(t-s,\) 
such that for all $X\in L^2((\Omega,\mathcal{F},\bP);\R^d)$
it holds that
\begin{align}
\Exp \left\| \Phi_{\delta}\left(X ,\frac{W_t - W_s}{\sqrt{t-s}}\right) \right\|^2
 \leq
 C^2 \Exp \| X \|^2.
\end{align}

\begin{theorem}
\label{thm:ChaosDecompNum}
Let 
$ J\in \mathbb{N} $. 
Let $ X_{\Delta, 0}\in L^2((\Omega,\mathcal{F}_0,\bP);\R^d) $,
and define $ (X_{\Delta, j\Delta})_{j=1}^{J} \in L^2(\Omega;(\R^{d})^J) $ by 
\begin{equation}\label{eq:defX}
    X_{\Delta, j\Delta} 
    = 
    \Phi_{\Delta}\left(
       X_{\Delta, (j-1)\Delta},  \frac{\Delta_{j} W}{\sqrt{\Delta}}
    \right)
\end{equation}
for all $ j \in \{1,2,\ldots,J\} $.
Let $ f\colon \R^d \rightarrow \R $ be measurable
and satisfy $ \Exp | f(X_{\Delta, T}) |^2 < \infty $.
Then 
\begin{eqnarray}
\label{wiener_chaos_simple}
  f(X_{\Delta, T})
   &=&
  \Exp[f(X_{\Delta, T}) | X_0 ]
  \\
  \nonumber
  && +
  \sum_{k = 1}^{\infty}
    \sum_{j=1}^{J}
      \sum_{i=1}^{m} 
        a_{k,j,i}(X_{\Delta, (j-1)\Delta},
      (\Delta_{j}W^r)_{r=1}^{i-1})H_{k}\left(\frac{\Delta_{j} W^i}{\sqrt{\Delta}} \right),
\end{eqnarray}
where the coefficients $a_{k,j,i}\colon\R^{d+i-1}\to\R$
in~\eqref{wiener_chaos_simple} 
are given by
\begin{equation}
\label{a_coeff}
  a_{k,j,i}(x,y)  =  \Exp\left[
      \left.
      f(X_{\Delta, T}) 
      H_k \left( \frac{\Delta_{j} W^i}{\sqrt{\Delta}}\right)
      \right| 
      X_{\Delta, (j-1)\Delta}=x,
      (\Delta_{j}W^r)_{r=1}^{i-1}=(y^r)_{r=1}^{i-1}
  \right]
\end{equation}
for all $k\in \mathbb{N}$, $j\in \{1,\ldots,J\}$ 
and $i\in \{1,\ldots,m\}$. 
\end{theorem}
\begin{rem}
The analogue of the main representation \eqref{wiener_chaos_simple} is of the form
\begin{gather*}
f(X_{\Delta,T})=\Exp[f(X_{\Delta,T})]+  
\sum_{j=1}^{J} \sum_{i=1}^m \sum_{1\leq q_1<\ldots<q_i\leq m} \sum_{k\in \mathbb{N}^i}  a_{k,j,i,q}(X_{\Delta, (j-1)\Delta})
\prod_{r=1}^i H_{k_r}\left(\frac{\Delta_{j} W^{q_r}}{\sqrt{\Delta}}\right),
\end{gather*}
with
\begin{eqnarray*}
a_{k,j,i,q}(x)=\mathrm{E}\left[\left. f(X_{\Delta,T}) \prod_{r=1}^i H_{k_r}(\Delta_{j} W^{q_r}/\sqrt{\Delta})\right|X_{\Delta, (j-1)\Delta}=x \right]
\end{eqnarray*}
and can be proved along the same lines
as~\eqref{wiener_chaos_simple}.
\end{rem}

Let us compare (in the one-dimensional case for the ease of notation) the representations \eqref{wiener_chaos_simple} and \eqref{wiener_chaos_l}.
First of all,  \eqref{wiener_chaos_simple}  has the form
\begin{gather*}
  f(X_{\Delta, T})
   =
  \Exp[f(X_{\Delta, T}) | X_0 ]
  +
  \sum_{k = 1}^{\infty}
    \sum_{j=1}^{J}
        a_{k,j}(X_{\Delta, (j-1)\Delta})H_{k}\left(\frac{\Delta_{j} W}{\sqrt{\Delta}} \right)
\end{gather*}
with
\begin{eqnarray*}
a_{k,j}(X_{\Delta, (j-1)\Delta})= 
    \sum_{
	  p\in I_{j,k}
	}
	  c_{p}(f(X_{\Delta, J\Delta}))
	  \left(
	    \prod_{\ell=0}^{j}
	      H_{p_{\ell}}(\Delta_{\ell} W / \sqrt{\Delta} )
	  \right).
\end{eqnarray*}
Denote 
\begin{equation}
\label{eq:fin:cv}
\mathcal{M}_{K,J}:=\sum_{k = 1}^{K}
    \sum_{j=1}^{J}
        a_{k,j}(X_{\Delta, (j-1)\Delta})H_{k}\left(\frac{\Delta_{j} W}{\sqrt{\Delta}} \right)
\end{equation}        
and 
\[
M_{K,J} := \sum_{k=1}^K \sum_{|p|=k} c_{p} \prod_{i=1}^{J} H_{p_i}(\Delta_{i} W/\sqrt{\Delta}).
\]
 The difference between the control variates  \(\mathcal{M}_{K,J}\) and \({M}_K\) can be written as
 $$
\mathcal{M}_{K,J} - M_{K,J}= \sum_{k\ge K+1}\sum_{j=1}^J\sum_{k'=1}^{K} \left[\sum_{|p|=k,\ p\in I_{j,k'}} c_{p} \prod_{i=1}^{J} H_{p_i}\left(\frac{\Delta_{i} W}{\sqrt{\Delta}}\right)\right]. 
$$
This implies that
$$\V(f(X_{\Delta,T})-\mathcal{M}_{K,J})\le \V(f(X_{\Delta,T})-M_{K,J}).$$
This latter inequality turns out to be very useful, as it is easier to analyze the truncation error related to the control variates  $M_{K,J}$ than the one connected to \(\mathcal{M}_{K,J}\).
We can write $X_{\Delta,j\Delta}=F(\sqrt{\Delta}\cdot\xi_1,\ldots,\sqrt{\Delta}\cdot\xi_j)$ for some function $F:\mathbb{R}^{j\cdot m}\to \mathbb{R}^{d},$ \(j=1,\ldots,J,\) where $\xi_j=\frac{\Delta_{j} W}{\sqrt{\Delta}},$ \(j=1,\ldots,J,\) are  $m$-dimensional Brownian increments and \(X_{\Delta,0}=x.\) This notation will be used in the next theorem, which assesses the efficiency of the control variate \(M_{K,\Delta}\).
\begin{theorem}
Consider the equation \eqref{x_sde} and its discretization given by 
$$
    X_{\Delta, j\Delta} 
    = 
    \Phi_{\Delta}\left(
       X_{\Delta, (j-1)\Delta},  \frac{\Delta_{j} W}{\sqrt{\Delta}}
    \right).$$
 Assume that function \(f(F(x, y_1,\ldots,y_J))\) is \(p\) times differentiable  in \(y\in \mathbb{R}^J\) such that 
 \begin{eqnarray*}
\Exp\left[\frac{\partial^{p_1+\ldots+p_J}f(F(\sqrt{\Delta}\cdot \xi_1,\ldots,\sqrt{\Delta}\cdot \xi_J))}{\partial \xi_1^{p_1}\dots\partial \xi_J^{p_J}}\right]
\end{eqnarray*}
is uniformly bounded in \(J\) (\(\Delta=T/J\)) and \(p\in \mathbb{N}^J\) with \(|p|\leq K.\)
Then 
$$\V(f(X_{\Delta,J\Delta})-M_{K,\Delta})\lesssim\Delta^{K}.$$
\end{theorem}
\begin{proof}
Due to the independence of Brownian increments and orthogonality of Hermite polynomials, we get
$$
\V\left[ \sum_{k \ge K+1 }^{\infty}
   \, \sum_{
      \substack{ 
	p\in \mathbb{N}_0^{(J+1) \times m}
	\\
		|p| = k
      }
    }
      c_{p}(f(X_{\Delta, J\Delta}))
      \prod_{j=0}^{J} 
	\prod_{i=1}^{m}
	  H_{p_{j,i}}\left(\frac{\Delta_{j} W^i}{\sqrt{\Delta}} \right)\right]=\sum_{k \ge K+1 }^{\infty}
   \, \sum_{
      \substack{ 
	p\in \mathbb{N}_0^{(J+1) \times m}
	\\
		|p| = k
      }
    } c^2_{p}.$$
Using the integration by parts in the case of one-dimensional diffusion (the multidimensional case is absolutely the same), we get for any $c_{p}$:
\begin{gather*}
\Exp\left[f(X_{\Delta,T})\times \prod_{i=1}^{J} H_{p_i}\left(\frac{\Delta_{i} W}{\sqrt{\Delta}}\right)\right]=\Exp\left[F(x, \sqrt{\Delta}\cdot\xi_1,\ldots,\sqrt{\Delta}\cdot\xi_J)\times \prod_{i=1}^{J} H_{p_i}\left(\xi_i\right)\right]
\\
=\frac{\Delta^{(K+1)/2}}{\prod_{i=1}^{J} p_i!}\Exp\left[\frac{\partial^{p_1+\ldots+p_J}f(F(x, \sqrt{\Delta}\cdot \xi_1,\ldots,\sqrt{\Delta}\cdot \xi_J))}{\partial \xi_1^{p_1}\dots\partial \xi_J^{p_J}}\right].
\end{gather*}
\end{proof}
\begin{remark}
Suppose  that \(f(x)\equiv x.\)  Then using the chain rule, we get for the Euler scheme:
\begin{eqnarray*}
\frac{\partial f(F(\sqrt{\Delta}\cdot \xi_1,\ldots,\sqrt{\Delta}\cdot \xi_J))}{\partial \xi_j}&=& 
\prod\limits_{i=j+1}^{J} \left(1+\frac{\partial \mu}{\partial x}(X_i)\,\Delta + \frac{\partial \sigma}{\partial x}(X_i)\sqrt{\Delta}\cdot \xi_i\right)\sigma(X_j).
\end{eqnarray*}
Taking expectation and using conditioning, we get
\begin{eqnarray}
\label{eq:rem_trunc}
\Exp\left[\frac{\partial f(F(\sqrt{\Delta}\cdot \xi_1,\ldots,\sqrt{\Delta}\cdot \xi_J))}{\partial \xi_j}\right]=\prod\limits_{i=j+1}^{J} \left(1+\Exp\left[\frac{\partial \mu}{\partial x}(X_i)\right]\,\Delta \right)\Exp[\sigma(X_j)].
\end{eqnarray}
 Hence the left hand side of \eqref{eq:rem_trunc} is uniformly bounded in \(J,\) provided 
 the expectations \(\Exp\left[\frac{\partial \mu}{\partial x}(X_i)\right],\) \(i=j+1,\ldots,J,\) and \(\Exp[\sigma(X_j)]\)  are bounded.
\end{remark}
\section{Regression approach to control variate construction}
In order to use the control variate \eqref{eq:fin:cv}
we need to compute the coefficients \label{a_coeff}.
 Since
\begin{equation}
  a_{k,j,i}(x,y)  =  \Exp\left[
      \left.
      f(X_{\Delta, T}) 
      H_k \left( \frac{\Delta_{j} W^i}{\sqrt{\Delta}}\right)
      \right| 
      X_{\Delta, (j-1)\Delta}=x,
      (\Delta_{j}W^r)_{r=1}^{i-1}=(y^r)_{r=1}^{i-1}
  \right]
\end{equation}
we can use nonparametric regression to estimate them, and therefore define coefficients $\mu_1$, $\mu_2$ and $\mu_3$. 
\subsection{General nonparametric approach}
To ease the explanation we now consider on $D+1$-dimensional random vector $(X,Y)$
where $X$ is~$\mathbb{R}^{D}$-valued and $Y$ is~$\mathbb{R}$-valued. Suppose that we want to find an approximation which is ``close to'' the $\mathbb{R}$-valued function 
\begin{align}
\label{regr:a}
a(x):=\mathrm{E}\left[Y|X=x\right].
\end{align}
Let us choose $Q$ real-valued functions $\psi_{1},\ldots,\psi_{Q}$ 
on $\mathbb{R}^{D}$ and simulate
a big number
$N$ of samples from the distributions of $X$ and $Y$.
In what follows these $N$ samples 
are denoted by $\mathcal{D}_{N}$:
$$
\mathcal{D}_{N}\doteq
\left\{
(X^{(n)},Y^{(n)}):
n=1,\ldots,N
\right\}.
$$
Let
$\beta=(
\beta_{1},\ldots,\beta_{Q})$ 
 be a solution of the following least squares optimisation problem:
\begin{align*}
\operatorname{argmin}_{\beta\in\mathbb{R}^{Q}}
\sum_{n=1}^{N}\left[Y^{(n)}-\sum_{k=1}^Q\beta_{k}\psi_{k}(X^{(n)})
\right]^{2}.
\end{align*}
Define an estimate for the 
function $a$ via
\begin{align*}
\hat a(x)\doteq
\hat a(x,\mathcal{D}_{N})\doteq
\sum_{k=1}^Q\beta_{k}\psi_{k}(x),
\quad x\in\mathbb{R}^{d}.
\end{align*}
The intermediate expression
$\hat a(x,\mathcal{D}_{N})$
in the above formula
emphasises that the estimates
$\hat a$
of the functions $a$
are random in that they depend on
the simulated samples.
The cost of computing
$\beta$ is of order $NQ^{2}$,
 since $\beta$
 is of the form $\beta=B^{-1}b$
 with 
\begin{align}
\label{regr:b_matr_reg}
B_{k,l}\doteq\frac{1}{N}\sum_{n=1}^{N}\psi_{k}\bigl(X^{(n)}\bigr)\psi_{l}\bigl(X^{(n)}\bigr)
\end{align}
and 
\begin{align*}
b_{k}\doteq\frac{1}{N}\sum_{n=1}^{N}\psi_{k}\bigl(X^{(n)}\bigr)\,Y^{(n)},
\end{align*}
where $k,l\in\{1,\ldots,Q\}$. 
In what follows,
we use the notation
$\bP_X$
for the distribution of $X$. In particular, we will work with the corresponding $L^2$-norm:
\begin{align*}
\|g\|^2_{L^2(\bP_X)}\doteq \int\limits_{\mathbb{R}^D} g^2(x)\,\bP_X(dx)=\mathbb{E}\left[g^2\left(X\right)\right].
\end{align*}
We assume that, 
for some positive
constants $\Sigma$ and $A$, it holds
\begin{enumerate}
\item[(A1) ] $\,\sup_{x\in\R^D}\V[Y|X=x]
\le\Sigma<\infty$,
\item[(A2) ] $\,\sup_{x\in\R^D} |a(x)|\le A<\infty$.
\end{enumerate}
Next we denote by $\tilde a$ the truncated regression estimate,
which is defined as follows: 
\begin{equation}
\label{regr:eq:30042016a1}
\tilde a(x)\doteq
T_{A}\hat a(x)
\doteq\begin{cases}
\hat a(x)&\text{if }
|\hat a(x)|\le A,\\
A\mathrm{sgn}(\hat a(x))
&\text{otherwise.}
\end{cases}
\end{equation}

We again emphasise that, in fact,
$\tilde a(x)=\tilde a(x,\mathcal{D}_{N})$,
that is, the estimates
$\tilde a$
of the functions $a$
depend on the simulated 
samples. 
Under (A1)--(A2) we obtain the following $L^2$-upper bound:
\begin{align}
\label{regr:eq:2104a2_mod}
\Exp\|\tilde a-a\|^2_{L^2(\bP_X)}
\le
\tilde c\left(\Sigma+A^2(\log N+1)\right)\frac{Q}{N}+8\inf_{g\in\Psi_{Q}}\| a-g\|^2_{L^2(\bP_X)},
\end{align}
where 
$
\Psi_{Q}\doteq\text{span}\left(\left\{\psi_{1},\ldots,\psi_{Q}\right\}\right)
$
and $\tilde c>0$ is a universal constant (cf.~Theorem~11.3 in~\cite{gyorfi2002distribution}).
Let us introduce the assumption that the function $a$ can be
well approximated by the functions from
$\Psi_{Q}$ in the sense that there are constants
$\kappa>0$ and $D_\kappa>0$ such that
\begin{align}
\label{regr:assump_int}
\inf_{g\in\Psi_{Q}}
\|a-g\|^2_{L^2(\bP_X)}
\leq  \frac{D_\kappa}{Q^\kappa}.
\end{align}
Note that this is a natural condition to be satisfied
for good choices of $\Psi_Q$.
So under assumptions (A1), (A2) and \eqref{regr:assump_int}, we get 

\begin{eqnarray*}
\Exp\|\tilde a-a\|^2_{L^2(\bP_X)}
\lesssim 
\frac{Q}{N}+\frac{1}{Q^\kappa}.
\end{eqnarray*}
Let us now consider the control variate 
\begin{eqnarray}
\label{control_variate_used}
  \tilde{\mathcal{M}}_{K,J}
   &=&
  \sum_{k = 1}^{K}
    \sum_{j=1}^{J}
      \sum_{i=1}^{m} 
        \tilde a_{k,j,i}(X_{\Delta, (j-1)\Delta},
      (\Delta_{j}W^r)_{r=1}^{i-1})H_{k}\left(\frac{\Delta_{j} W^i}{\sqrt{\Delta}} \right),
\end{eqnarray}
where \(\tilde a_{k,j,i}\) are estimated using nonparametric regression with \(Q\) basis functions. It's easy to see, that 
$$\V(f(X_{\Delta,J\Delta})-\tilde{\mathcal{M}}_{K,J})\preceq \Delta^K + \Delta^{-1}\cdot\left(\frac{Q}{N}+\frac{1}{Q^\kappa}\right)$$
under the corresponding assumptions on the functions  \(a_{k,j,i}.\)
At this point it is very important to emphasize, that we still have $$\Exp [\tilde{\mathcal{M}}_{K,J}|\mathcal{D}_{N}]=0,$$
which means, that at no point we introduce additional bias due to the finite number of basis functions or due to the numerical discretization. 
Now taking into account that the number of coefficients to compute is of order \(J K^d NQ^2,\) we can take for any 
fixed \(K>1,\) \(\mu_3=K,\) \(\mu_2=(K+1)/\kappa+1,\) \(\mu_1=(K+1)(1+3/\kappa)+1\) to get for the overall cost of the variance reduced MLMC algorithm 
$$\mathcal{C}^\varepsilon_{\mu_1,\mu_2,\mu_3,\beta}= \varepsilon^{-\max\left(2-\frac{2\eta}{\eta+\mu_1},\frac{1}{\alpha}\right)},\quad \varepsilon\to 0.$$
 with $\eta=\min(\beta-1,K(1-1/\kappa)-1/\kappa-1).$ 
 So the complexity order is alway below $\varepsilon^{-2},$ provided   \(\beta>1,\) \(K>(\kappa+1)/(\kappa-1),\) \(\kappa>1\) and \(\alpha>1/2.\)

\section{Numerical experiments}
We consider the SDE
\begin{align}
\notag
dX_t^i&=-\sin\left(X_t^i\right)\cos^3\left(X_t^i\right)dt+\cos^2\left(X_t^i\right)dW_t^i,\quad X_0^i=0,\quad i\in\left\{1,2,3,4\right\},\\
\label{5d_sde}
dX_t^5&=\sum_{i=1}^4\left[-\frac{1}{2}\sin\left(X_t^i\right)\cos^2\left(X_t^i\right)dt+\cos\left(X_t^i\right)
dW_t^i\right]+dW_t^5,\quad X_0^5=0.
\end{align}
The solution of~\eqref{5d_sde} is given by
$$
X_t^i=\arctan\left(W_t^i\right),\ i\in\left\{1,2,3,4\right\},
X_t^5=\sum_{i=1}^4\operatorname{arsinh}\left(W_t^i\right)
+W_t^5.
$$
for $t\in\left[0,1\right]$. Further, we consider the functional 
\begin{align*}
f(x)=\cos\left(\sum_{i=1}^5x^i\right)-20\sum_{i=1}^4\sin\left(x^i\right),
\end{align*}
that is, we have
\begin{align*}
\mathbb{E}\left[f\left(X_1\right)\right]&=\left(\mathbb{E}\left[\cos\left(
\arctan\left(W_1^1\right)+\operatorname{arsinh}\left(W_1^1\right)\right)\right]\right)^4
\mathbb{E}\left[\cos\left(W_1^5\right)\right]
\end{align*}
We use the an antithetic MLMC approach from \cite{gs14}, where the following Milstein alike discretization scheme has been utilized:
\begin{gather*}
X_{n+1}^i=X_{n}^i-\sin\left(X_{n}^i\right)\cos^3\left(X_{n}^i\right)\Delta+\cos^2\left(X_{n}^i\right)\Delta W^i_n -\cos^3(X_{n}^i)\cdot\sin(X_{n}^i)\cdot\left(\Delta^2 W^i_n-\Delta\right)\\
X_{n+1}^{5}=\sum_{i=1}^4\left[-\frac{1}{2}\sin\left(X_{n}^i\right)\cos^2\left(X_{n}^i\right)\Delta+\cos\left(X_t^i\right)\Delta W^i_n + \frac12\cos\left(X_t^i\right)\sin\left(X_t^i\right)\cdot\left(\Delta^2 W^i_n-\Delta\right)\right]+\Delta W^5_n
\end{gather*}
for $i=1,2,3,4$ and $X^i_0=X^5_0=0$. We use the updated Antithetic MLMC estimator
\begin{gather*}
\widehat Y\doteq\frac{1}{n_{L_0}}\sum_{i=1}^{n_{L_0}}\bigl[f(X^{(i)}_{L_0,T})-M^{(i)}_{2,2^{L_0}}\bigr]+\sum\limits _{l=L_0+1}^{L}Y_l=\\\frac{1}{n_{L_0}}\sum_{i=1}^{n_{L_0}}\bigl[f(X^{(i)}_{L_0,T})-M^{(i)}_{2,2^{L_0}}\bigr]+\sum\limits _{l=L_0+1}^{L}\frac{1}{n_{l}}\sum_{i=1}^{n_{l}}\bigl[\frac12\left(f(X^{(i)}_{f,l,T})+f(X^{(i)}_{a,l,T})\right)-f(X^{(i)}_{{l-1},T})\bigr],
\end{gather*}
where subindices stand for discretized paths with antithetic approach, discussed in details in \cite{gs14}, which we refer to due to the length constraints. This scheme with coupling recovers variance decay rate $\beta=2$. 
We consider accuracies $\log_2(\varepsilon)=-4,-7, -10, -13$, and we set $L_0=1,2,3,4$ respectively, which corresponds to our approach to use polynomial partitioning with $p=3$.
\subsection{Regression for control variate}
\begin{figure}[h!]
    \label{fig:mlmc:variance}
        \centering\includegraphics
        [width=\textwidth]{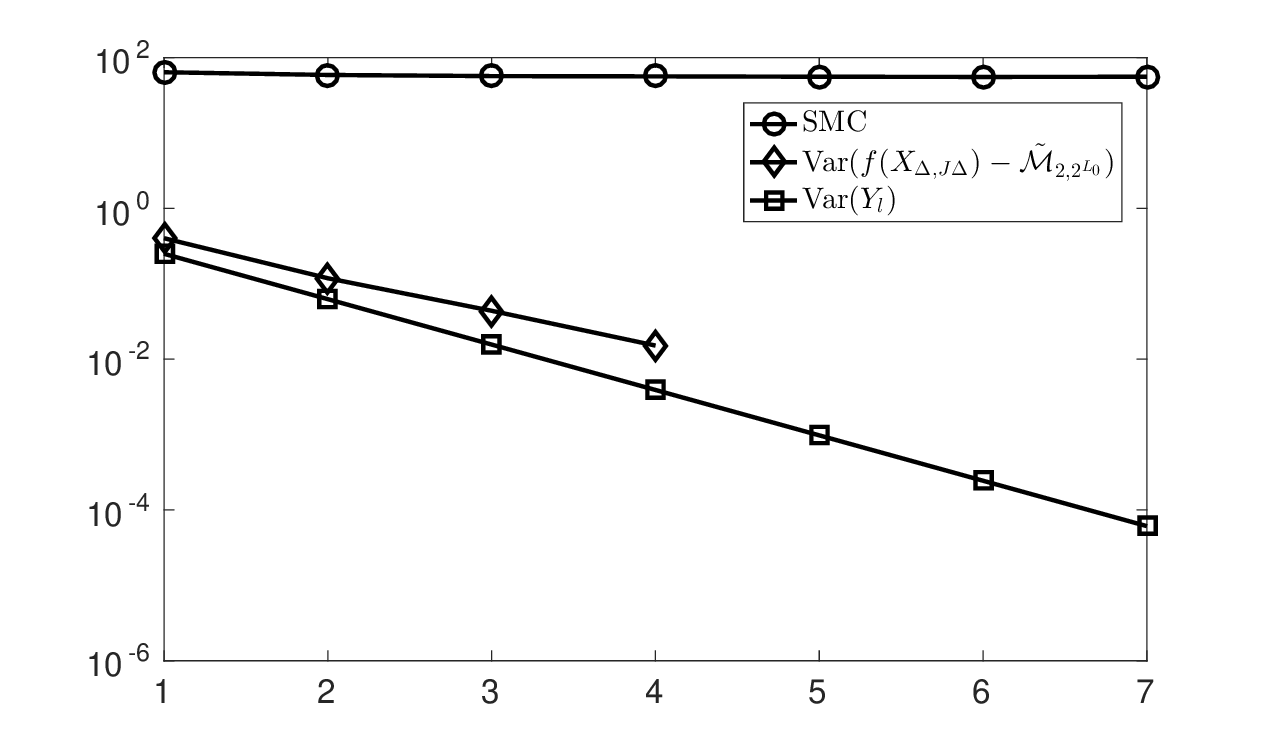}
        \caption{Variance decay of Antithetic MLMC (square markers), Single level MC (circle markers), and Single level with control variate $\tilde{\mathcal{M}}_{2,2^{L_0}}$ (diamond markers).}
    \end{figure}

\begin{figure}[h!]
        \label{cv:mlmc:cost}
        \centering\includegraphics
        [width=\textwidth]{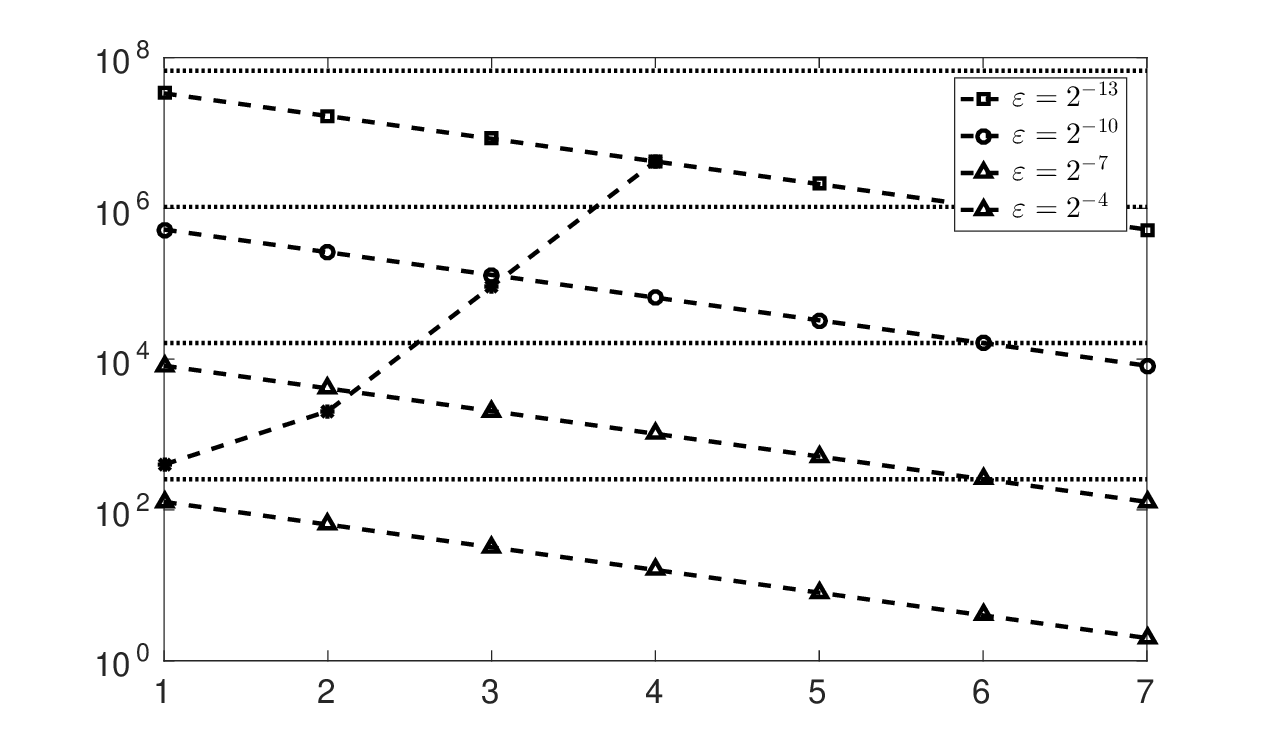}
        \caption{Dotted straight lines correspond to the expected MLMC complexity of order $\varepsilon^{-2}$ for different values of $\varepsilon$. Decaying lines correspond to MLMC complexities from level of the form $\varepsilon^{-2}\cdot 2^{-(\beta-1)}=\varepsilon^{-2}\cdot 2^{-l}$. The cost of $\tilde{\mathcal{M}}_{2,2^{L_0}}$ construction is presented on a dashed line with star-shaped markers.}
    \end{figure}

We consider the control variate \eqref{control_variate_used}, with $K=2$ and we want to find an estimator for it $\tilde{\mathcal{M}}_{2,2^{L_0}}$. It's easy to see, that 
$$\V(f(X_{\Delta,J\Delta})-\tilde{\mathcal{M}}_{2,2^{L_0}})\preceq \Delta_{L_0}^2 + \Delta_{L_0}^{-1}\cdot\left(\frac{Q}{N}+\frac{1}{Q^\kappa}\right)$$ 
that all the drift, diffusion and the functional are sufficiently regular, so in the case of polynomial regression of order $p$, we will have $\kappa=2\cdot p$.  Taking into account that the overall regression cost is of order $\Delta^{-1} NQ^2$, we set 
$$Q\asymp \Delta^{-\frac{3}{\kappa}},\ N\asymp \Delta^{-3-\frac{3}{\kappa}}\Rightarrow \mu_1=4+\frac{9}{\kappa},\ \mu_2=1+\frac{3}{\kappa},\ \mu_3=2.$$ 
In our numerical experiments we will focus on piecewise polynomial approximation of order $p=3$, hence $\kappa=6$, which leads to 
$$\eta=\min(\beta-1,K(1-1/\kappa)-1/\kappa-1)=\min(1,2(1-1/6)-1/6-1)=0.5.$$ 
$$
\mathcal{C}^\varepsilon_{\mu_1,\mu_2,\mu_3,\beta}=\varepsilon^{-1\tfrac{5}{6}},\quad \Delta_{L_0}\asymp \varepsilon^{\tfrac{1}{3}}.
$$
The results, describing the effectiveness of the control variate construction and the variance decay of the antithetic MLMC can be seen on Figure \ref{fig:mlmc:variance}. Due to sufficient regularity of drift, diffusion and functional (moreover, we work here with the bounded functionals), our numerical results are able to reproduce perfectly our expectations from the theory. The variance for MLMC and Single level MC methods are estimated based on $2.5\cdot 10^6$ paths, while the variance of Single level MC is estimated on $10^6$ paths, with the control variates constructed from $N=\max(\Delta_l^{-3.5},100)$ paths.
    The complexity of $\tilde{\mathcal{M}}_{2,2^{L_0}}$ estimation increases with $L_0$, so we need so set it accurately. According to our parameters choice we set $\Delta_{L_0}\asymp \varepsilon^{\tfrac{1}{3}},$ so for accuracies $\log_2(\varepsilon)=-4,-7, -10, -13$ we have $L_0=1,2,3,4$ respectively. The cost of constructing the control variates in our simulation has the form
    $$\Delta_{L_0}^{-1}\cdot N\cdot Q^2 = \Delta_{L_0}^{-1}\cdot \max(\Delta_l^{-3.5},100)\cdot \Delta_{L_0}^{-1},$$
    which is presented on Figure \ref{cv:mlmc:cost}. 
    There one can see reference dotted straight lines, which correspond to the expected Antithetic MLMC complexity of order $\varepsilon^{-2}$ for different values of $\varepsilon$. As an illustration of accuracy of our chosen parameters, we also plot the cost of estimating $\sum\limits _{l=L_0+1}^{L}Y_l$, which in the case of $\beta=2$ is proportional to $\varepsilon^{-2}\cdot 2^{-L_0}$. According to the complexity Theorem \ref{mlmc:cv:cost:thm}, $L_0$ corresponds to a level, when the overall cost of construction of control variate $\tilde{\mathcal{M}}_{2,2^{L_0}}$ (and the cost of simulating level $L_0$ along with the cost of sampling $\tilde{\mathcal{M}}_{2,2^{L_0}}$) is proportional to the cost of simulating all forthcoming Antithetic MLMC levels. As one can see on Figure \ref{cv:mlmc:cost}, our constraint on number of training paths $N$ doesn't satisfy this rule of thumb for $\varepsilon=2^{-4}$, while for other accuracies it matches almost perfectly. Moreover, the gain of Antithetic MLMC approach with control variate is clearly represented by the difference between the dotted straight lines for each accuracy and the level, where line of constructing the control variate cost intersects the decaying line of leftover MLMC levels calculation cost.
 \section{Conclusion}
 We presented a MLMC method with control variate on its starting level, which allows us to have a randomised Monte Carlo algorithm with complexity of order less than $\varepsilon^{-2}$. This approach is implementable in an arbitrary dimension, which is its strong advantage. Our analysis also suggests that other possible improvements for MLMC can be made, if one can improve the cost of the initial level in MLMC approach, as its cost  without variance reduction (determined by central limit theorem) prevent us from improving $\varepsilon^{-2}$ complexity bound, regardless of how good is the coupling. The disadvantage of the proposed nonparametric regression algorithm, but not the general control variate approach, is that it requires regularity of the problem and a sophisticated choice of basis functions. Both of these issues should be treated for considered problem specifically, but our numerical results suggest, that even very simple regression approach can give substantial saving, provided that the problem of interest is regular enough.
\section{Proofs}
\subsection{Proof of Theorem~\ref{thm:ChaosDecompNum}} 

Without loss of generality we may assume there exists a 
Gaussian vector, suggestively denoted by  
$ \Delta_{0}W \colon \Omega \rightarrow \R^m$
with a covariance operator $\Delta I_{\R^d}$,
such that $\mathcal{F}_0 = \sigma(\Delta_{0}W)$.
For $p\in \mathbb{N}_0^{m\times(J+1)}$ and 
$
  X 
  \in 
  L^2((\Omega,\mathcal{F},\bP),\R^d)
$
define 
\begin{align}\label{eq:coeff_Hermite}
  c_{p}(X)
  = 
  \Exp \left[
     X
     \prod_{j=0}^{J} 
	\prod_{i=1}^{m}
	    H_{p_{i,j}}( \Delta_{j} W^i/\sqrt{\Delta} )
  \right].
\end{align}
For $j\in \{0,1,\ldots,J\}$,
$i\in \{1,\ldots,m\}$,
and $k\in \mathbb{N},$
define
\begin{align}
 I_{j,i,k} 
 =
 \left\{ 
  p\in \mathbb{N}_0^{(J+1)\times m}
  \colon
  \begin{array}{l}
    p_{j,i} = k
    \text{ and }
    (
      \forall
      r \in \{i+1,\ldots,m\}
      \colon
      p_{j,r} = 0
    )
  \\
    \text{and }
    ( 
      \forall \ell \in \{j+1,\ldots, J \},
      \forall r\in \{1,\ldots,m\}
      \colon
      p_{\ell,r} = 0
    )
  \end{array}
  \right\}. 
\end{align}
The Wiener chaos expansion of $f(X_{\Delta, J\Delta})$ 
with respect to $(\Delta_{j}W)_{j=0}^{J}$
is given by
\begin{align}
\notag
  f(X_{\Delta, J\Delta})
  &
  =
  \Exp[ f(X_{\Delta, J\Delta}) ]
  +
  \sum_{k = 1 }^{\infty}
   \, \sum_{
      \substack{ 
	p\in \mathbb{N}_0^{(J+1) \times m}
	\\
		|p| = k
      }
    }\,
      c_{p}(f(X_{\Delta, J\Delta}))
      \prod_{j=0}^{J} 
	\prod_{i=1}^{m}
	  H_{p_{j,i}}( \Delta_{j} W^i/\sqrt{\Delta} )\\
	  \notag & 
  =
  \Exp[ f(X_{\Delta, J\Delta}) ]
  +
  \sum_{j=0}^{J}
    \sum_{i=1}^{m}
      \sum_{k=1}^{\infty}
	\sum_{
	  p\in I_{j,i,k}
	}
	  c_{p}(f(X_{\Delta, J\Delta}))
	  H_{k} ( \frac{\Delta_{j} W^i}{\sqrt{\Delta}} )\\
\label{eq:chaos_exp_f}
 & \qquad \qquad \times 
  \left(
    \prod_{\ell=0}^{j-1}
    \prod_{r=1}^{m}
      H_{p_{\ell,r}}( \Delta_{\ell} W^r /\sqrt{\Delta}  )
  \right)
  \left(
    \prod_{r=1}^{i-1}
      H_{p_{j,r}}( \Delta_{j} W^r/\sqrt{\Delta} )
  \right)
\end{align}
Then it follows that 
\begin{align}
  &\Exp[f(X_{\Delta, J\Delta}) | X_0]
   =
  \Exp[f(X_{\Delta, J\Delta}) | \Delta_{0}W ]=
\\
\notag &
  \Exp[f(X_{\Delta, J\Delta})]
  +
  \sum_{i=1}^{m}
      \sum_{k=1}^{\infty}
	\sum_{
	  p\in I_{0,i,k}
	}
	  c_{p}(f(X_{\Delta, J\Delta}))	  
	  H_{k} \left( \frac{\Delta_{0} W^i}{\sqrt{\Delta}} \right)
 \times
	  \left(
	    \prod_{r=1}^{m}
	      H_{p_{0,r}}\left(\frac{\Delta_{0} W^r}{\sqrt{\Delta}} \right)
	  \right),
\end{align}
and for all $j\in \{1,\ldots,J\}$,
all $i\in \{1,\ldots,m\}$, and 
all $k\in \mathbb{N}$ it holds that
\begin{align}
\label{eq:arep}
&  \Exp\left[
	\left.
	f(X_{\Delta, J\Delta}) 
	H_k ( \Delta_{j} W^i/\sqrt{\Delta} )
	\right| 
	(\Delta_{\ell}W)_{\ell=1}^{j-1},
	(\Delta_{j}W^r)_{r=1}^{i-1}
    \right]
\\
\notag =& 
  \Exp\left[
	\left.
	f(X_{\Delta, J\Delta}) 
	H_k ( \Delta_{j} W^i/\sqrt{\Delta} )
	\right| 
	X_{\Delta, (j-1)\Delta},
	(\Delta_{j}W^r)_{r=1}^{i-1}
    \right]
\\
\notag =& 
      \sum_{
	  p\in I_{j,i,k}
	}
	  c_{p}(f(X_{\Delta, J\Delta}))
	  \left(
	    \prod_{\ell=0}^{j-1}
	    \prod_{r=1}^{m}
	      H_{p_{\ell,r}}(\Delta_{\ell} W^r / \sqrt{\Delta} )
	  \right)
	  \left(
	    \prod_{r=1}^{i-1}
	      H_{p_{j,r}}( \Delta_{j} W^r/\sqrt{\Delta} )
	  \right).    
\end{align}
We can use now \eqref{eq:arep} to rewrite~\eqref{eq:chaos_exp_f} as \eqref{wiener_chaos_simple}.

\bibliographystyle{spmpsci}
\bibliography{ml_vr_bibliography}

\end{document}